\pdfoutput=1

\documentclass[runningheads]{llncs}
\usepackage{array}
\usepackage{multirow}
\usepackage{amsfonts}
\usepackage{mathtools}
\usepackage{subfig}
\usepackage{graphicx}
\usepackage[pscoord]{eso-pic}
\usepackage{textcomp}
\usepackage{hyperref}
\newcommand{\placetextbox}[3]{
  \setbox0=\hbox{#3}
  \AddToShipoutPictureFG*{
    \put(\LenToUnit{#1\paperwidth},\LenToUnit{#2\paperheight}){\vtop{{\null}\makebox[0pt][c]{#3}}}%
  }%
}%

\begin{document}

\placetextbox{0.5}{0.96}{\normalfont \small Author accepted manuscript, published in ``Modeling Decisions for Artificial Intelligence (MDAI 2020). Lecture Notes in Computer Science,}

\placetextbox{0.5}{0.945}{\normalfont \small vol 12256, 66--77, 2020. Springer, Cham.}

\placetextbox{0.5}{0.930}{\normalfont \small The final authenticated publication is available online at \url{https://doi.org/10.1007/978-3-030-57524-3\_6}.}

\title{An unsupervised capacity identification approach based on Sobol' indices\thanks{This work was supported by the São Paulo Research Foundation (FAPESP, grant numbers 2016/21571-4 and 2017/23879-9) and the National Council for Scientific and Technological Development (CNPq, grant number 311357/2017-2).}}
%
%
\author{Guilherme Dean Pelegrina\inst{1,3}\orcidID{0000-0001-7301-6167} \and
Leonardo Tomazeli Duarte\inst{2}\orcidID{0000-0003-0290-0080} \and
Michel Grabisch\inst{3} \and
Jo\~{a}o Marcos Travassos Romano\inst{1}}
\authorrunning{G. D. Pelegrina, L. T. Duarte, M. Grabisch and J. M. T. Romano}
%
\institute{School of Electrical and Computer Engineering, University of Campinas, 400 Albert Einstein Avenue, 13083-852 Campinas, Brazil \\
\email{pelegrina@decom.fee.unicamp.br}, \email{romano@dmo.fee.unicamp.br} \and
School of Applied Sciences, University of Campinas, 1300 Pedro Zaccaria Street, 13484-350 Limeira, Brazil \\
\email{leonardo.duarte@fca.unicamp.br} \and
Centre d'Économie de la Sorbonne, Université Paris I Panthéon-Sorbonne, 106-112 Boulevard de l'Hôpital, 75647 Paris Cedex 13, France \\
\email{michel.grabisch@univ-paris1.fr}}
\maketitle              
\begin{abstract}

In many ranking problems, some particular aspects of the addressed situation should be taken into account in the aggregation process. An example is the presence of correlations between criteria, which may introduce bias in the derived ranking. In these cases, aggregation functions based on a capacity may be used to overcome this inconvenience, such as the Choquet integral or the multilinear model. The adoption of such strategies requires a stage to estimate the parameters of these aggregation operators. This task may be difficult in situations in which we do not have either further information about these parameters or preferences given by the decision maker. Therefore, the aim of this paper is to deal with such situations through an unsupervised approach for capacity identification based on the multilinear model. Our goal is to estimate a capacity that can mitigate the bias introduced by correlations in the decision data and, therefore, to provide a fairer result. The viability of our proposal is attested by numerical experiments with synthetic data.


\keywords{Multicriteria decision making \and Multilinear model \and Unsupervised capacity identification \and Sobol' index.}
\end{abstract}
\section{Introduction}
\label{sec:intro}

In multicriteria decision making (MCDM)~\cite{Figueira2005}, a typical problem consists in obtaining a ranking of a set of alternatives (candidates, projects, cars, ...) based on their evaluations in a set of decision criteria. These evaluations are generally aggregated in order to achieve overall values for the alternatives and, therefore, to define the ranking. In the literature~\cite{Grabisch2009}, one may find several aggregation functions that can be used to deal with such problems. A simple example is the weighted arithmetic mean (WAM), which comprises a linear aggregation and is based on parameters representing weight factors associated to each criterion. Although largely used, there are some characteristics about the addressed decision problem that the WAM cannot deal with. An example is the interaction among criteria, which should be modelled in order to overcome biased results originated from the correlation structure of the decision data~\cite{Marichal2000}.

Different aggregation functions have been developed to model interactions among criteria. For instance, one may cite the well-known Choquet integral~\cite{Choquet1954,Grabisch1996}, which derives the overall evaluations through a piecewise linear function. Moreover, although less used in comparison with the Choquet integral, one also may consider the multilinear model~\cite{Owen1972}. In this case, we aggregate the set of evaluations through a polynomial function.

A drawback of both Choquet integral and multilinear model, in comparison with the WAM, is that we need many more parameters (the capacity coefficients) to model the interactions among criteria. Therefore, the task of capacity identification is an important issue to be addressed when considering these functions. In the literature, one may find some supervised approaches (i.e., based on the information about both criteria evaluations and overall values to be used as learning data) for both Choquet integral~\cite{Grabisch2008} and multilinear model~\cite{Pelegrina2020}. Moreover, one also may find unsupervised approaches (i.e., based only on the information about the decision data) to estimate the parameters of the Choquet integral~\cite{Duarte2018}.

Since in the unsupervised approaches one does not have access to the overall evaluations as learning data, one needs to assume a characteristic about the decision problem that we would like to deal with. This characteristic will be considered when implementing the capacity identification model. For instance,~\cite{Duarte2018} associates some Choquet integral parameters to similarity measures of pairs of criteria in order to deal with the bias provided by correlations in the decision data. Therefore, the goal is the estimation of a capacity that leads to fairer overall evaluations in the sense that this bias is mitigated.

Motivated by the interesting results obtained by~\cite{Duarte2018}, in this paper, we tackle the unsupervised capacity identification problem in the context of the multilinear model, which remains largely unknown in the literature. However, instead of using the similarity measures, we deal with correlations by considering the Sobol' indices of coalitions of criteria, which can be directly associated with the multilinear model~\cite{Grabisch2017}. In order to attest the efficacy of the proposal, we apply our approach in scenarios with different numbers of alternatives and different degrees of correlation.

The rest of this paper is organized as follows. Section~\ref{sec:theor} describes the underlying theoretical concepts of this paper. In Section~\ref{sec:method}, we present the proposed unsupervised approach to deal with the problem of capacity identification. Numerical experiments are conducted in Section~\ref{sec:exper}. Finally, in Section~\ref{sec:conclu}, we present our conclusions and future perspectives.

\section{Theoretical background}
\label{sec:theor}

This section presents the theoretical aspects associated with our proposal, mainly the multilinear model and the Sobol' indices.

\subsection{Multicriteria decision making and the multilinear model}
\label{sec:mcda}

The MCDM problem addressed in this paper comprises the ranking of $n$ alternatives $a_1, a_2, \ldots, a_n$ based on their evaluations with respect to a set $C$ of $m$ criteria. Generally, we represent the decision data in a matrix $\mathbf{V}$, defined by
\begin{equation}
\label{eq:dec_data}
\mathbf{V}=\left[\begin{array}{cccc}
v_{1,1} & v_{1,2} & \ldots & v_{1,m} \\
v_{2,1} & v_{2,2} & \ldots & v_{2,m} \\ 
\vdots & \vdots & \ddots & \vdots \\
v_{n,1} & v_{n,2} & \ldots & v_{n,m}
\end{array}
\right],
\end{equation}
where $v_{i,j}$ is the evaluation of alternative $a_i$ with respect to the criterion $j$. Therefore, in order to obtain the ranking, for each alternative $a_i$ we aggregate the criteria evaluations through an aggregation function $F(\cdot)$ and order the alternatives based on the overall evaluations $r_i = F(v_{i,1}, \ldots, v_{i,m})$.

As mentioned in Section~\ref{sec:intro}, candidates for $F(\cdot)$ are the WAM and the multilinear model. The WAM is defined as
\begin{equation}
\label{eq:model_wam}
F_{WAM}(v_{i,1}, \ldots, v_{i,m}) = \sum_{j=1}^m w_j v_{i,j},
\end{equation}
where $w_j$ ($w_j \geq 0$, for all $j=1, \ldots, m$, and $\sum_j^m w_j = 1$) represents the weight factor associated to criterion $j$. On the other hand, the multilinear model~\cite{Owen1972} is defined as
\begin{equation}
\label{eq:model_ml}
F_{ML}(v_{i,1}, \ldots, v_{i,m}) = \sum_{A \subseteq C} \mu(A) \prod_{j \in A} v_{i,j} \prod_{j \in \overline{A}} \left(1-v_{i,j} \right),
\end{equation}
where $v_{i,j} \in \left[0,1 \right]$, $\overline{A}$ is the complement set of $A$ and the parameters $\mu = \left[\mu(\emptyset), \mu(\left\{ 1\right\}), \ldots, \mu(\left\{ m\right\}), \mu(\left\{ 1,2\right\}), \ldots, \mu(\left\{ m-1,m\right\}), \ldots, \mu(C)\right]$, called capacity~\cite{Choquet1954}, is a set function $\mu:2^{C} \rightarrow \mathbb{R}$ satisfying the following axioms\footnote{It is worth mentioning that the multilinear model generalizes the WAM, i.e., if we consider an additive capacity, $F_{ML}(\cdot)$ is equivalent to $F_{WAM}(\cdot)$.}:

\begin{itemize}
\item $\mu(\emptyset) = 0$ and $\mu(C) = 1$ (boundedness),

\item for all $A \subseteq B \subseteq C$, $\mu(A) \leq \mu(B) \leq \mu(C)$ (monotonicity).
\end{itemize}

Let us illustrate the application of the considered aggregation functions in the problem of ranking a set of students based on their grades in a set of subjects (we adapted this example from~\cite{Grabisch1996}). The decision data as well as the overall evaluations and the ranking positions for both WAM and multilinear model are described in Table~\ref{tab:example}. For instance, we consider that $w_1=w_2=w_3=1/3$ and $\mu = \left[0, 1/3, 1/3, 1/3, 2/5, 2/3, 2/3, 1\right]$.

\begin{table}[ht]
  \begin{center}
  \caption{Illustrative example.}\label{tab:example}
  {
  \renewcommand{\arraystretch}{1.2}
	\small
  \begin{tabular}{cccccccccc}
	\cline{1-4}\cline{6-7}\cline{9-10}
		\multirow{2}{*}{\textbf{Students}} & \multicolumn{3}{c}{\textbf{Grades}} &  &  \multicolumn{2}{c}{\textbf{WAM}} &  & \multicolumn{2}{c}{\textbf{Multilinear model}} \\
		\cline{2-4}\cline{6-7}\cline{9-10}
     & Mathematics & Physics & Literature &  & $r_i$ & Position &  & $r_i$ & Position \\
		\cline{1-4}\cline{6-7}\cline{9-10}
		Student 1 & 1.00 & 0.94 & 0.67 &  & 0.8700 & 1 &  & 0.7874 & 2 \\
		Student 2 & 0.67 & 0.72 & 0.94 &  & 0.7767 & 3 &  & 0.7703 & 3 \\
		Student 3 & 0.83 & 0.89 & 0.83 &  & 0.8500 & 2 &  & 0.8172 & 1 \\
		\cline{1-4}\cline{6-7}\cline{9-10}
  \end{tabular}
  }
  \end{center}
\end{table}

One may note that student 1 has an excellent performance on both mathematics and physics, but a lower grade in literature in comparison with the other students. Student 2 has the lowest grades in mathematics and physics and a very good one in literature. Finally, student 3 has an equilibrated performance, with good grades in all disciplines. With respect to the ranking, by applying the WAM, student 1 achieves the first position. However, one may consider that both mathematics and physics are disciplines that are associated with the same latent factor, i.e., they are correlated. Therefore, an aggregation that takes into account this interaction may lead to fairer results by avoiding this bias. That is the case of the ranking provided by the multilinear model, in which the student 3, the one with equilibrated grades, achieves the first position.

\subsection{Interaction indices and 2-additive capacity}

In the previous section, we defined the multilinear model in terms of a capacity $\mu$. However, the capacity coefficients do not have a clear interpretation. Therefore, we normally use an alternative representation of $\mu$, called interaction index~\cite{Grabisch1997}. In the context of the multilinear model, the Banzhaf interaction index~\cite{Roubens1996} is suitable (see, e.g.,~\cite{Pelegrina2020}), and is defined as follows:
\begin{equation}
\label{eq:mutoii}
I^\mathcal{B}(A) = \frac{1}{2^{\left| C \right|-\left| A \right|}}\sum_{D \subseteq C\backslash A} \sum_{D' \subseteq A} \left(-1\right)^{\left| A \right| - \left| D' \right|}\mu(D' \cup D), \, \forall A \subseteq C,
\end{equation}
where $\left| A \right|$ represents the cardinality of the subset $A$. If we consider, for example, a singleton $j$, we obtain the Banzhaf power index $\phi^\mathcal{B}_j \in [0,1]$, given by
\begin{equation}
\label{eq:power_ind}
\phi^\mathcal{B}_j = \frac{1}{2^{\left| C \right|-1}}\sum_{D \subseteq C\backslash \left\{j\right\}} \left[\mu(D \cup \left\{j\right\}) - \mu(D) \right],
\end{equation}
which can be interpreted as the marginal contribution of criterion $i$ alone taking into account all coalitions. Moreover, if we consider a pair of criteria $j,j'$, we obtain the interaction index $I^\mathcal{B}_{j,j'}$ expressed by
\begin{equation}
\label{eq:int_pair_ind}
I^\mathcal{B}_{j,j'} = \frac{1}{2^{\left| C \right|-2}}\sum_{D \subseteq C\backslash \left\{j,j'\right\}} \left[\mu(D \cup \left\{j,j'\right\}) - \mu(D \cup \left\{j\right\}) - \mu(D \cup \left\{j'\right\}) + \mu(D)\right].
\end{equation}
In this case, the interpretation is the following:
\begin{itemize}
\item if $I^\mathcal{B}_{j,j'} < 0$, there is a negative interaction between criteria $j,j'$, which models a redundant effect between them,
\item if $I^\mathcal{B}_{j,j'} > 0$, there is a positive interaction between criteria $j,j'$, which models a complementary effect between them,
\item if $I^\mathcal{B}_{j,j'} = 0$, there is no interaction between criteria $j,j'$, which means that they act independently.
\end{itemize}

In the multilinear model used to deal with the example presented in the last section, the interaction indices associated with the considered $\mu$ are $I^\mathcal{B} = \left[0.4675, 0.2667, 0.2667, 0.4017, -0.1367, 0.1333, 0.1333, 0.2600\right]$. Therefore, we have a negative interaction between criteria 1 and 2, positive interactions between the others pairs of criteria and a higher power index for criterion 3 in comparison to the other ones. It is worth mentioning that, given $I^\mathcal{B}(A)$, $\forall A \subseteq C$, one may retrieve $\mu(A)$ through the following transform
\begin{equation}
\label{eq:iitomu}
\mu(A) = \sum_{D \subseteq C} \left(\frac{1}{2}\right)^{\left| D \right|}\left(-1\right)^{\left|D\backslash A \right|}I^\mathcal{B}(D), \, \forall A \subseteq C.
\end{equation}

By using either the capacity $\mu$ or the interaction index $I^\mathcal{B}$, we have $2^m-2$ parameters to be determined in order to use the multilinear model. This may pose a problem in some situations, since the number of parameters increases exponentially with the number of criteria. In that respect, one may adopt a specific capacity, called 2-additive~\cite{Grabisch1997}, which reduces the number of unknown parameters to $m(m+1)/2 - 1$. We say that a capacity $\mu$ is 2-additive if the interaction index $I^\mathcal{B}(A) = 0$ for all $A$ such that $\left| A \right| \geq 3$. In this paper, we consider such a capacity in the multilinear model.

\subsection{Sobol' indices}

In several applications it is useful to analyse the sensitivity of a model output given a subset of all input factors. For this purpose, it is usual to carry out a decomposition of the model into terms associated with different input variables. For instance, consider the high-dimensional model representation (HDMR) of a model $Y = f(Z_1, Z_2, \ldots, Z_m)$ given by
\begin{equation}
\label{eq:decomp}
Y = f_{\emptyset} + \sum_{j=1}^m f_j(Z_j) + \sum_{j < j'}^m f_{j,j'}(Z_j,Z_{j'}) + \ldots + f_C(Z_C),
\end{equation}
where $f_{\emptyset}, f_j, f_{j,j'}, \ldots, f_C$, $\forall j,j' \subseteq C$, are terms of increasing dimensions. A possible function $f$ that can be used is the following:


\begin{equation}
f_A(Z_A) = \sum_{D \subseteq A} (-1)^{|A \backslash D |} \mathbb{E}[Y|Z_D],
\end{equation}
where $\mathbb{E}[Y|Z_D]$ denotes the conditional expectation of $Y$ given the variables $Z_j$, for all $j \in D$. In particular, $f_{\emptyset} = \mathbb{E}[Y]$, $f_j(Z_j) = \mathbb{E}[Y|Z_j] - f_{\emptyset}$ and $f_{j,j'}(Z_j,Z_{j'}) = \mathbb{E}[Y|Z_j, Z_{j'}] - f_j(Z_j) - f_j(Z_j) - f_{\emptyset}$.

As mentioned in~\cite{Saltelli2008}, one may estimate $\mathbb{E}[Y|Z_j]$ by cutting the $Z_j$ domain into slices and calculating the average value for each slice. Therefore, if these values have a pattern, $\mathbb{E}[Y|Z_j]$ has a large variation across $Z_j$, which means that this variable is ``important'' (or has a high ``impact'') in the output model. On the other hand, if no pattern is found, the variation across $Z_j$ is small and this input is ``less important'' (or has a ``less impact'') in the model output.

By using the function $f$, one may calculate the variation across a variable through the variance of $f_j(Z_j)$, i.e., $\text{Var}[f_j(Z_j)] = \text{Var}[\mathbb{E}[Y|Z_j]]$. For instance, if we consider a single variable $Z_j$ and normalize it, one obtains the first-order Sobol' sensitivity index, given by
\begin{equation}
\label{eq:sobol_first}
S_j = \frac{\text{Var}[\mathbb{E}[Y|Z_j]]}{\text{Var}[Y]},
\end{equation}
which is a measure that indicates the degree of the aforementioned ``importance'' (or the ``impact'') that the input variable $Z_j$ has in the model output. The same conclusions can be obtained for higher-order terms, i.e., for any coalition $A \subseteq C$, 
\begin{equation}
\label{eq:sobol_all}
S_A = \frac{\text{Var}[f_A(Z_A)]}{\text{Var}[Y]}
\end{equation}
represents a measure of the ``importance'' (or the ``impact'') that the coalition of variables $Z_A$ have in the model output. Under the assumption that the input variables (criteria, in the addressed decision problem) are independent and that they follow a uniform distribution on $[0,1]$,~\cite{Grabisch2017} presented an interesting result:

\begin{theorem}{(Grabisch and Labreuche~\cite{Grabisch2017})}
\label{theo:sobol}
Consider the multilinear model $F_{ML}$ of a capacity $\mu$. The (nonnormalized) Sobol’ index of a subset $\emptyset \neq A \subseteq C$ is given by
\begin{equation}
\text{Var}[(F_{ML})_A] = \frac{1}{3^{\left| A \right|}}\left(\hat{\mu}(A)\right)^2,
\end{equation}
where $\hat{\mu}$, defined by
\begin{equation}
\hat{\mu}(A) = \left(\frac{-1}{2}\right)^{\left| A \right|}I^{\mathcal{B}}(A),
\end{equation}
is the Fourier transform of $\mu$.
\end{theorem}

Therefore, one may remark that the Sobol' index is associated with the Banzhaf interaction index through the equation
\begin{equation}
\label{eq:sobol_banz}
\text{Var}[(F_{ML})_A] = \frac{1}{12^{\left| A \right|}}\left(I^{\mathcal{B}}(A)\right)^2.
\end{equation}
This relation will be used in our proposal to deal with the unsupervised capacity identification problem.

\section{The proposed unsupervised approach for capacity identification}
\label{sec:method}

\begin{figure}[ht]
\centering
\subfloat[Criteria 1 and 2.]{\includegraphics[width=1.79in]{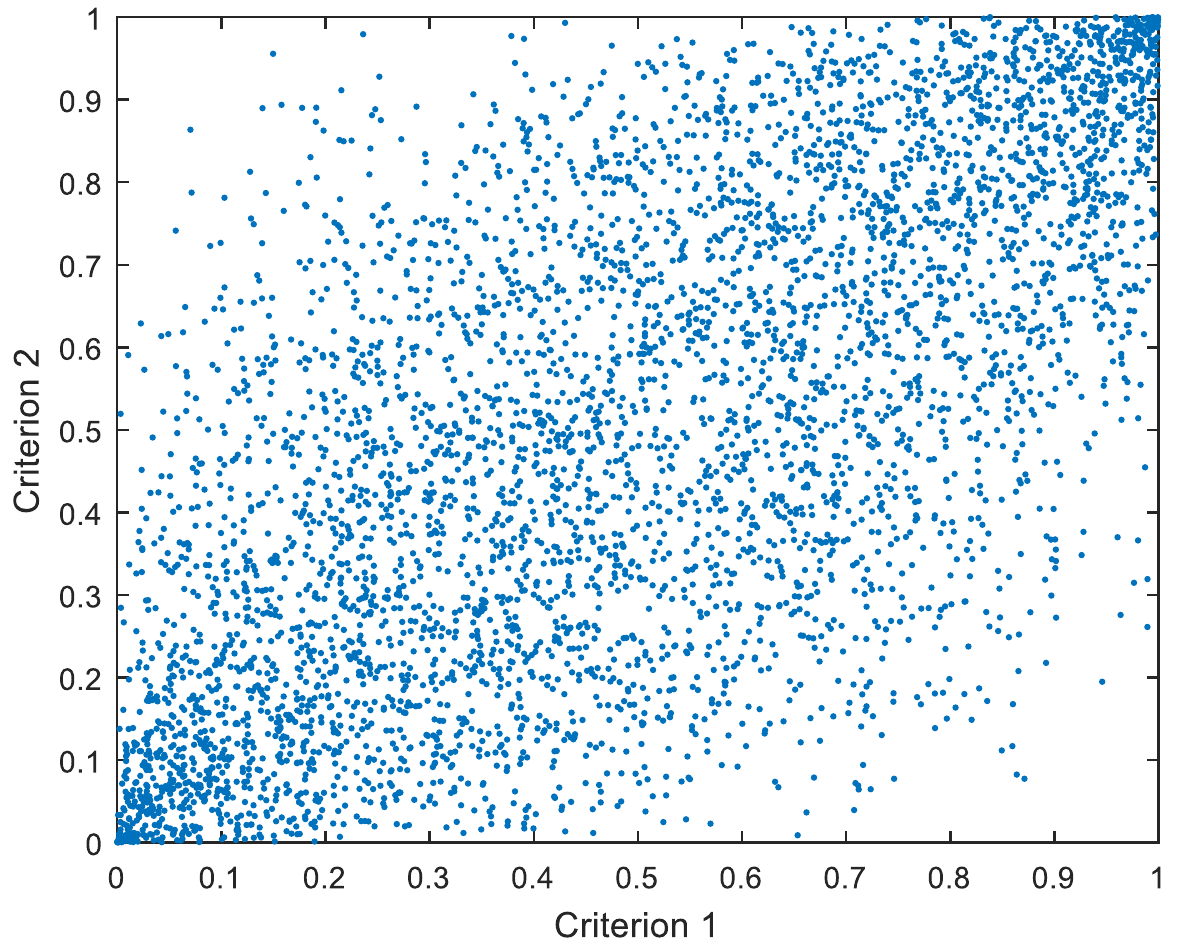}
\label{fig:scatter12}}
\hfil
\subfloat[Criteria 1 and 3.]{\includegraphics[width=1.79in]{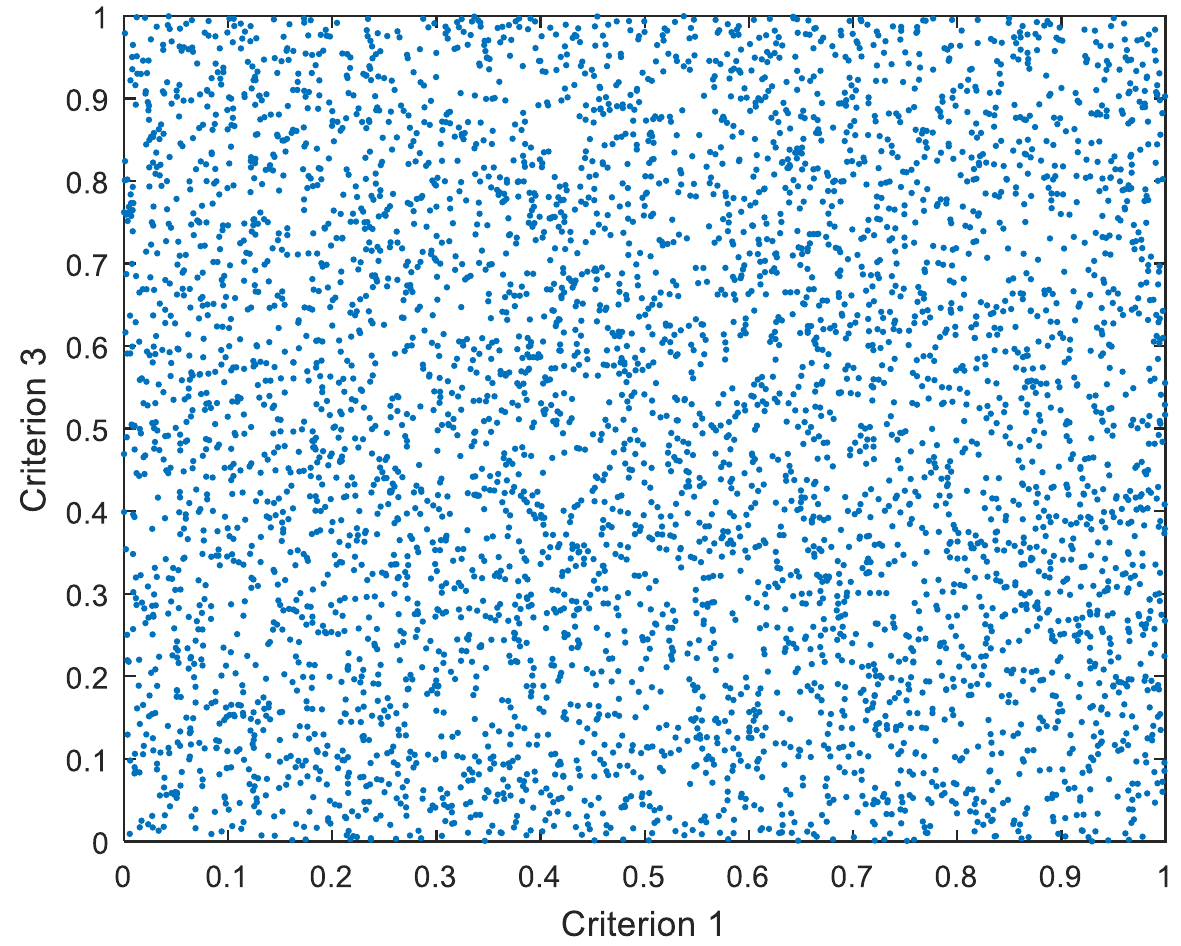}
\label{fig:scatter13}}
\hfil
\subfloat[Criteria 2 and 3.]{\includegraphics[width=1.79in]{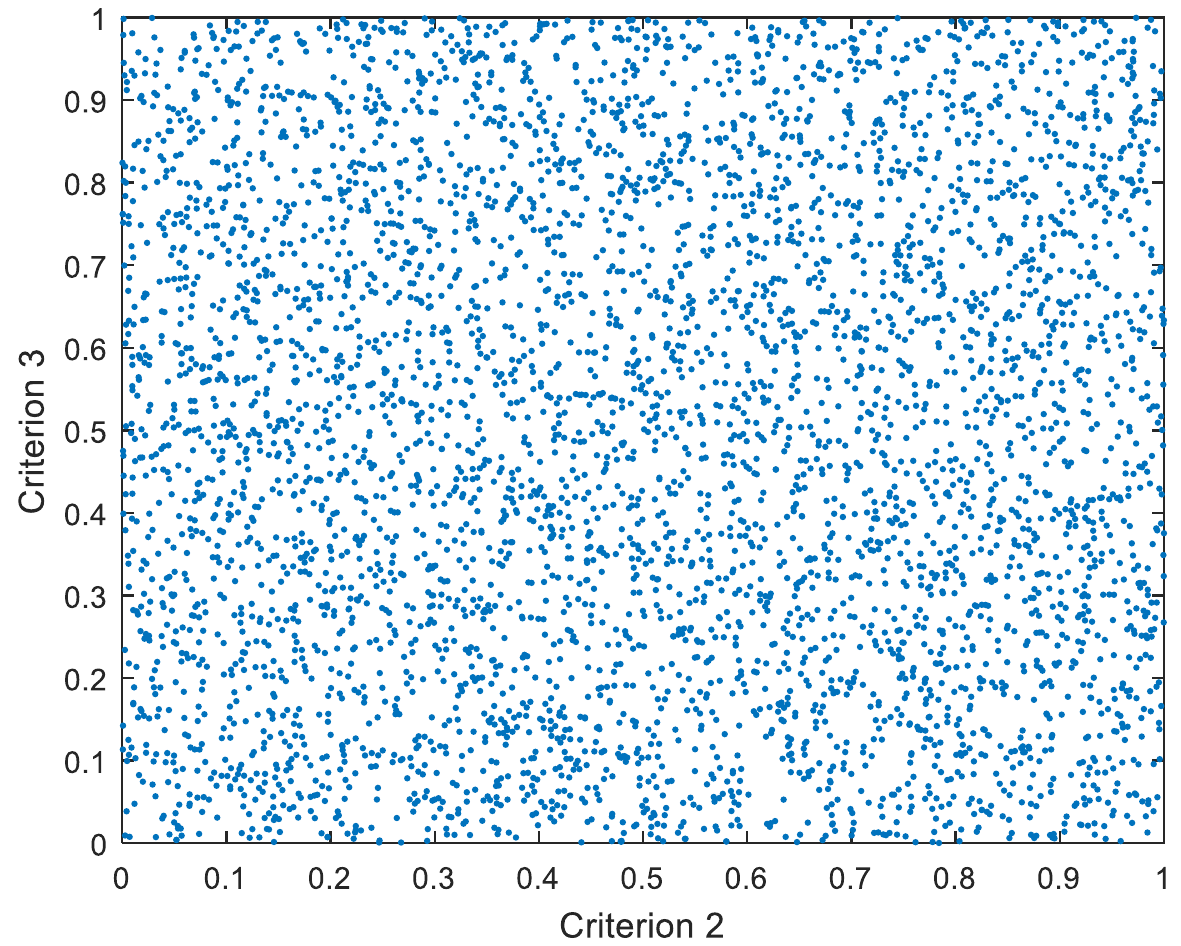}
\label{fig:scatter23}}
\caption{Scatter plot of pairs of criteria.}
\label{fig:scatter}
\end{figure}

In order to present the motivations for our proposal, let us consider a decision problem composed by 5000 alternatives and 3 criteria (generated according to a uniform distribution on $[0,1]$). For instance, this situation may comprise the problem of ranking the set of students described in Section~\ref{sec:mcda}. Figure~\ref{fig:scatter} presents the scatter plot among pairs of criteria. One may note that criteria 1 and 2 are correlated (with Pearson correlation coefficient $\rho_{1,2} = 0.6768$).

Suppose that we want to analyse the impact that the criteria 1 and 3 alone have in the output model. Moreover, assume an additive capacity, i.e., $\mu = [0, 1/3, 1/3, 1/3, 2/3, 2/3, 2/3, 1]$ (in this case, $\phi^\mathcal{B}_1 = \phi^\mathcal{B}_2 = \phi^\mathcal{B}_3 = 1/3$). If we calculate the nonnormalized Sobol' index according to Equation~\eqref{eq:sobol_banz}, one obtains $\text{Var}[(F_{ML})_1)] = \text{Var}[(F_{ML})_3)] = (1/12)\left(\phi^\mathcal{B}_3\right)^2 \approx 0.0093$. On the other hand, if we consider Equation~\eqref{eq:sobol_first} (with $Y$ being the multilinear model and without normalization), one obtains $\text{Var}[\mathbb{E}[F_{ML}|\mathbf{v}_1]] \approx 0.0263$ and $\text{Var}[\mathbb{E}[F_{ML}|\mathbf{v}_3]] \approx 0.0091$, where $\mathbf{v}_j = \left[v_{1,j}, v_{2,j}, \ldots, v_{n,j} \right]$. Therefore, one may remark that we achieved similar values for both $\text{Var}[(F_{ML})_3)]$ and $\text{Var}[\mathbb{E}[F_{ML}|\mathbf{v}_3]]$, but very different ones comparing $\text{Var}[(F_{ML})_1)]$ and $\text{Var}[\mathbb{E}[F_{ML}|\mathbf{v}_1]]$. This difference is due to the existing correlation between criteria 1 and 2, which violates the hypothesis about the independence between the input variables assumed by Theorem~\ref{theo:sobol}. As a consequence, since these criteria are positively correlated, $\text{Var}[\mathbb{E}[F_{ML}|\mathbf{v}_1]]$ is also influenced by criterion 2, which increases the impact of criterion 1 in the output model. With respect to $\text{Var}[\mathbb{E}[F_{ML}|\mathbf{v}_3]]$, since criterion 3 has no correlation with any other criterion, its value remains independent on the other inputs.

Based on this discussion, our hypothesis is that the difference that we achieved with respect to the Sobol' indices may be reduced by applying a capacity that takes into account interactions among criteria. Suppose a scenario in which we have neither further information about the capacity coefficients nor overall evaluations provided by the decision maker. In this situation, it may be interesting to adopt a capacity that is able to compensate the bias provided by the correlations in the decision data and lead to fairer overall evaluations for the alternatives. In this context, this can be achieved by a capacity that leads to a multilinear model in which all coalitions of criteria with the same cardinality have similar impacts on the obtained overall evaluations. Therefore, the aim in this paper is to adjust a capacity $\mu$ in order to minimize the difference between the Sobol' indices for subsets of criteria with the same cardinality. Mathematically, the optimization problem is given by\footnote{It is worth mentioning that we must satisfy the axioms of a capacity.}
\begin{equation}
\label{eq:optmodel2}
\begin{array}{ll}
\displaystyle\min_{\mathbf{\mu}} & \displaystyle\sum_{\substack{A \subset C, \\ A \neq \emptyset}}\sum_{\substack{D \subset C, \\ \left| D \right| = \left| A \right|}}\left(S_A - S_D \right)^2.
\end{array}
\end{equation}

\section{Numerical experiments}
\label{sec:exper}

This section presents the numerical experiments and the obtained results.

\subsection{Application of our proposal in the illustrative example}

As a first experiment, let us apply the proposed approach in the decision problem addressed in Section~\ref{sec:method}. In order to reduce the number of parameters to be estimated, but keeping a flexibility to model interactions, we considered the 2-additive multilinear model and a capacity $\mu$ such that $\mu(\left\{ 1\right\})=\mu(\left\{ 2\right\})=\mu(\left\{ 3\right\})=1/3$ (i.e., in the absence of further information about the capacity coefficients, we predefine the same value for all $\mu(\left\{ j\right\})$). Therefore, one also need to estimate the capacity coefficients associated with pairs of criteria.

With respect to the optimization problem, other than the axioms of capacity presented in Section~\ref{sec:mcda}, we must also guarantee that $I^\mathcal{B}(C) = 0$, which leads to the following condition:
\begin{equation}
\label{eq:cond_opt}
\mu(\left\{ 1,2\right\})+\mu(\left\{ 1,3\right\})+\mu(\left\{ 2,3\right\})=2.
\end{equation}
Moreover, aiming at achieving an aggregation function whose individual criteria have similar impacts on the overall evaluations, we only considered the minimization of the difference between first-order Sobol' indices, i.e., the subsets $A$ in~\eqref{eq:optmodel2} are restricted to singletons $j$. In order to solve the optimization problem, we adopted a simple iterative heuristic method based on the golden section search~\cite{Vajda1989}. For instance, we started with the additive capacity and selected at random a $\mu(\left\{ j,j'\right\})$ to be fixed ($\mu(\left\{ 1,3\right\}) = 2/3$, for example). Thereafter, we selected another $\mu(\left\{ j'',j'''\right\})$ ($\mu(\left\{ 1,2\right\})$, for example) to be optimized and applied the golden section search in the associated dimension to deal with the addressed optimization problem. In other words, by fixing $\mu(\left\{ j,j'\right\})$, we perform a one-dimensional search on $\mu(\left\{ j'',j'''\right\})$ that solves~\eqref{eq:optmodel2}. Since we must satisfy~\eqref{eq:cond_opt}, one may note that, when applying the golden section search on $\mu(\left\{ j'',j'''\right\})$, the other capacity coefficient that was not selected so far is automatically adjusted ($\mu(\left\{ 2,3\right\}) = 2 - 2/3 - \mu(\left\{ 1,2\right\})$, in this case). This procedure is repeated until the convergence to the minimum (possibly a local minimum).

Based on the aforementioned assumptions, the application of the proposed approach in the illustrative example led to the capacity and the associated interaction indices presented in Table~\ref{tab:example_result}.

\begin{table}[ht]
  \begin{center}
  \caption{Achieve capacity and interaction indices.}\label{tab:example_result}
  {
  \renewcommand{\arraystretch}{1.2}
	\small
  \begin{tabular}{ccccccccc}
	\cline{2-9}
		 & \multicolumn{8}{c}{$A$} \\
		\cline{2-9}
     & $\emptyset$ & $\left\{ 1\right\}$ & $\left\{ 2\right\}$ & $\left\{ 3\right\}$ & $\left\{ 1,2\right\}$ & $\left\{ 1,3\right\}$ & $\left\{ 2,3\right\}$ & $C$ \\
		\hline
		\multicolumn{1}{c|}{$\mu(A)$} & 0 & 0.3333 & 0.3333 & 0.3333 & 0.4190 & 0.7778 & 0.8033 & 1 \\
		\multicolumn{1}{c|}{$I^{\mathcal{B}}(A)$} & 0.5000 & 0.2650 & 0.2778 & 0.4572 & -0.2477 & 0.1111 & 0.1366 & 0 \\
		\hline
  \end{tabular}
  }
  \end{center}
\end{table}

One may note that we achieved $I^{\mathcal{B}}_{1,2} < 0$, which was expected since both criteria 1 and 2 are correlated. Moreover, the obtained Banzhaf power index $\phi^\mathcal{B}_3$ was higher in comparison to the other ones, which also contributes to increase the impact of criterion 3 (the independent one) in the output model. With respect to the first-order Sobol' indices, we achieved $S_1 \approx S_2 \approx S_3 \approx 0.0173$.

If we consider the problem of ranking the students, which can be configured as the considered decision problem, the estimated capacity leads to the overall evaluations $r_1=0.7976$, $r_2=0.8196$ and $r_3=0.8445$. Therefore, the student 3 will also be the first one in the ranking.

\subsection{Experiments varying the number of alternatives and the degree of the correlation}

\begin{figure}[h!]
\centering
\includegraphics[height=5.0cm]{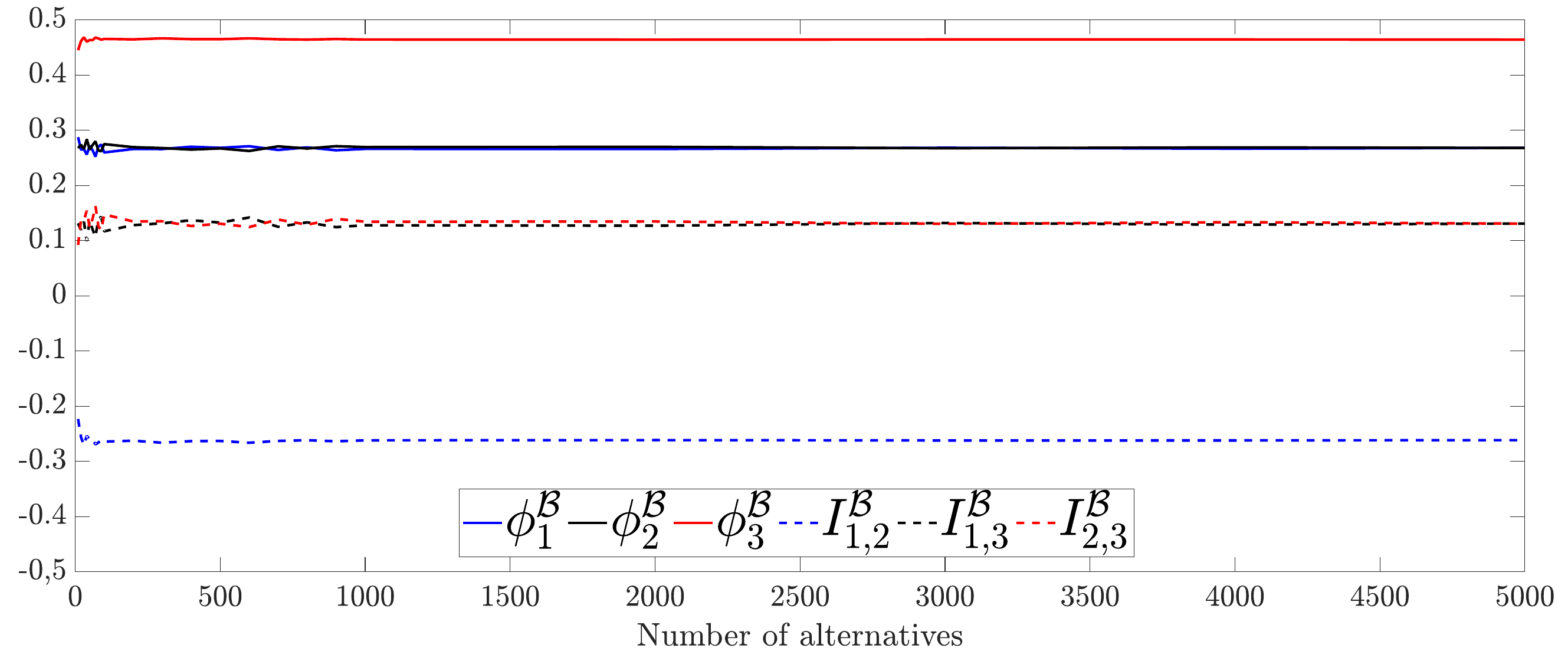}
\caption{Results for $\rho_{1,2} \approx 0.75$.}
\label{fig:result_p075}
\end{figure}

\begin{figure}[h!]
\centering
\includegraphics[height=5.0cm]{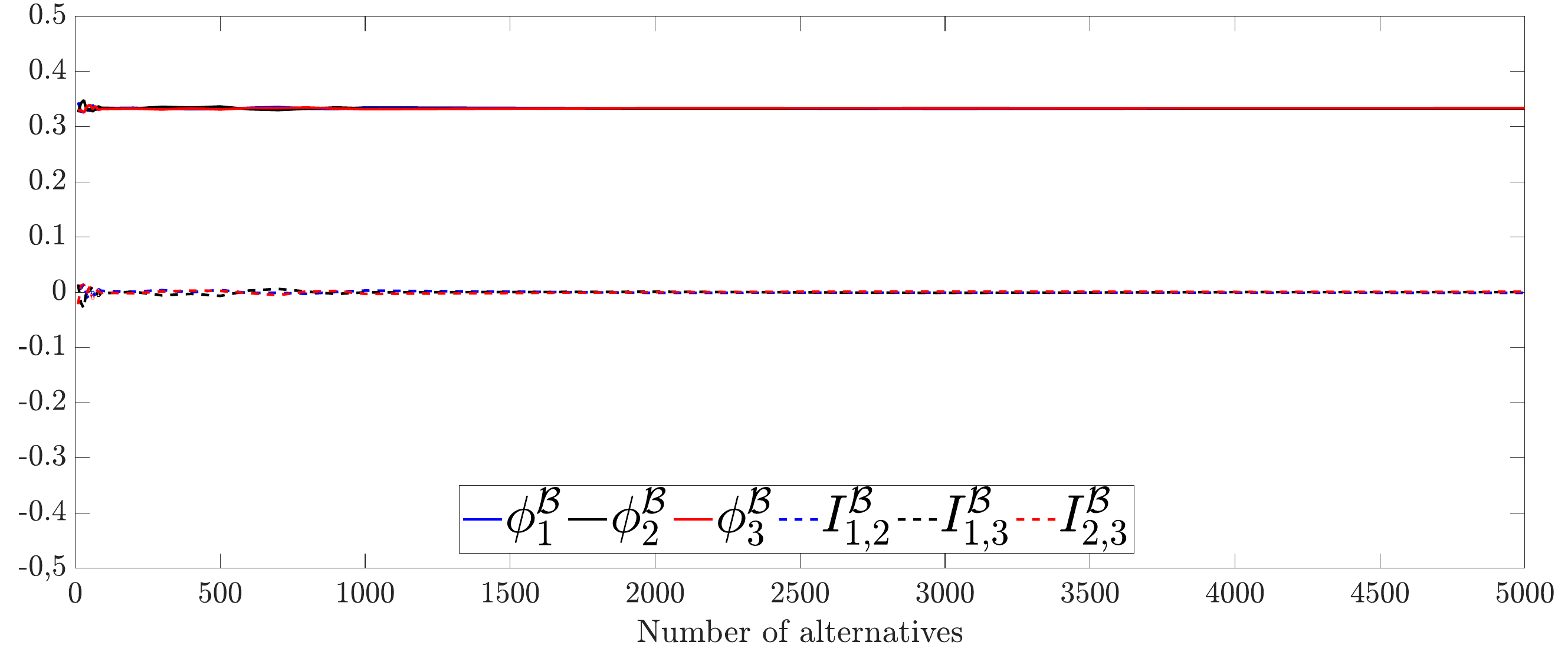}
\caption{Results for $\rho_{1,2} \approx 0$.}
\label{fig:result_pn0}
\end{figure}

\begin{figure}[h!]
\centering
\includegraphics[height=5.0cm]{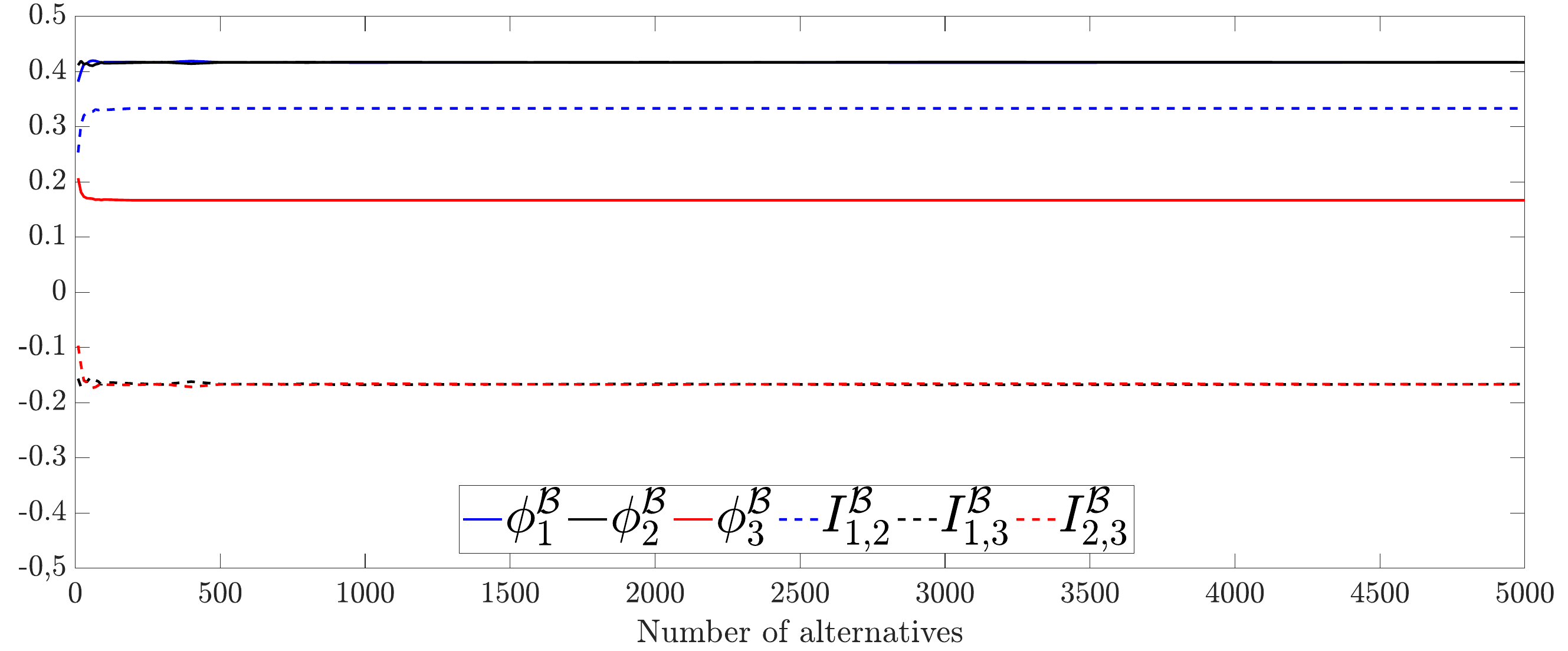}
\caption{Results for $\rho_{1,2} \approx -0.75$.}
\label{fig:result_n075}
\end{figure}

In order to further investigate our proposal, we considered several different scenarios, varying the number of alternatives and the degree of the correlation between criteria 1 and 2. In all cases we considered 3 decision criteria and generated the evaluations according to a uniform distribution on $[0,1]$. Based on the same assumptions considered in the last experiment, the obtained interaction indices (averaged over 100 simulations) for decision problems with $\rho_{1,2} \approx 0.75$, $\rho_{1,2} \approx 0$ and $\rho_{1,2} \approx -0.75$ are presented in Figures~\ref{fig:result_p075},~\ref{fig:result_pn0} and~\ref{fig:result_n075}, respectively.

One may note that, in all cases, the stability in the obtained capacity increases as the number of alternatives also increases. This is due to the statistics involved in the Sobol' indices calculation, which require more data to be well estimated.

In Figure~\ref{fig:result_p075}, we clearly see that, in order to compensate the positive correlation between criteria 1 and 2, we achieved $I^{\mathcal{B}}_{1,2} < 0$ (redundant effect for the correlated criteria), both $I^{\mathcal{B}}_{1,3}$ and $I^{\mathcal{B}}_{2,3} > 0$ (positive interaction for the independent criteria) and a marginal contribution of criterion 3 ($\phi^{\mathcal{B}}_3$) greater than the other ones. Conversely, in Figure~\ref{fig:result_n075}, the negative correlation between criteria 1 and 2 led to $I^{\mathcal{B}}_{1,2} > 0$ (complementary effect for the correlated criteria), both $I^{\mathcal{B}}_{1,3}$ and $I^{\mathcal{B}}_{2,3} < 0$ (negative interaction for the independent criteria) and a marginal contribution of criterion 3 ($\phi^{\mathcal{B}}_3$) lower than the other ones.

With respect to Figure~\ref{fig:result_pn0}, which contains the results when all the criteria are independent, one may see that the obtained capacity is an additive one. Therefore, we do not need to model interactions or increase marginal contributions to equilibrate the Sobol' indices, since they are already similar.

\section{Conclusions}
\label{sec:conclu}

It is usual to observe the presence of correlations between criteria in multicriteria decision making problems. In these cases, the obtained ranking may be biased towards alternatives that have good evaluations in correlated criteria, i.e., that measure the same latent factor. Even with a worst performance in the other criteria, these alternatives can achieve better positions compared to the ones whose evaluations are more equilibrated.

In order to deal with these situations by modelling interactions among criteria, one may use aggregation functions such as the Choquet integral or the multilinear model. These functions are based on a capacity, i.e., a set of parameters associated with all possible coalitions of criteria. Therefore, the number of parameters increases with the number of criteria, making it difficult to define or estimate these values.

In this paper, we addressed the problem of capacity identification in an unsupervised fashion. Differently form our previous work~\cite{Pelegrina2020}, we do not consider any further information about the parameters or overall values provided by the decision maker. Our approach aims at extracting information contained in the decision data and estimating a capacity that can compensate the correlations among criteria. For instance, we assumed that all singletons should have the same impact on the output model and used the Sobol' indices as a means of comparison.

The obtained results attested the application of the proposed approach. In situations with positive (resp. negative) correlation between a pair of criteria, the achieved associated interaction index was negative (resp. positive), which models a redundancy (resp. complementarity) effect. Moreover, the power indices were also adjusted in order to balance the Sobol' indices.

For future perspectives, we would like to analyse the heuristic used to deal with the optimization problem. Other assumptions about the capacities as well as different search algorithms may be investigated. Moreover, we here addressed the situation in which the criteria are not independent. Therefore, future works can be conducted to verify the impact that distributions different from the uniform may have in the Sobol' indices. Finally, we also intend to apply the proposal in real datasets.

%
%

%
%
%
\bibliographystyle{splncs04}
\bibliography{_refs_mdai2020}

\end{document}